\documentclass[12pt,superscriptaddress]{revtex4}
\usepackage{graphicx}
\usepackage{amsmath}
\usepackage{amssymb}
\usepackage{bm}

\begin{document}

\title{Forecasting Financial Extremes: A Network Degree Measure of Super-exponential Growth}

\author{Wanfeng Yan}
 \email{wanfeng.yan@gmail.com}
 \affiliation{Banque Pictet \& Cie SA, Route des Acacias 60, 1211 Geneva 73, Switzerland} %

\author{Edgar van Tuyll van Serooskerken}
 \affiliation{Banque Pictet \& Cie SA, Route des Acacias 60, 1211 Geneva 73, Switzerland} %

\begin{abstract}
Investors in stock market are usually greedy during bull markets and scared during bear markets. The greed or fear spreads across investors quickly. This is known as the herding effect, and often leads to a fast movement of stock prices. During such market regimes, stock prices change at a super-exponential rate and are normally followed by a trend reversal that corrects the previous over reaction. In this paper, we construct an indicator to measure the magnitude of the super-exponential growth of stock prices, by measuring the degree of the price network, generated from the price time series. Twelve major international stock indices have been investigated. Error diagram tests show that this new indicator has strong predictive power for financial extremes, both peaks and troughs. By varying the parameters used to construct the error diagram, we show the predictive power is very robust. The new indicator has a better performance than the LPPL pattern recognition indicator.

\end{abstract}
\maketitle \clearpage


\section{Introduction}

Many models are designed to describe the dynamics of financial bubbles and crashes, among which the log-periodic power law (LPPL) (also known as the Johansen-Ledoit-Sornette (JLS)) model \cite{js,jsl,jls,sornettecrash} stating that bubbles are generated by behaviors of investors and traders that create positive feedback in the valuation of assets. Therefore, bubbles are not characterized by an exponential increase of prices but rather by a faster-than-exponential growth of prices. This unsustainable growth pace often ends with a finite-time singularity at some future time $t_c$. After $t_c$, the old regime with a faster-than-exponential growth rate is broken, and often but not always, the new regime is a dramatic fall over a short period, a financial crash. Yan et.~al \cite{rebound} extended the JLS model by introducing the concept of ``negative bubbles'' as the mirror image of standard financial bubbles. They found that positive feedback mechanisms also exist in negative bubbles, which are in general associated with large rebounds or rallies. Therefore, before financial extremes, like crashes and rebounds, a super-exponential growth in price is often observed. Measuring super-exponential growth in financial time series can be used to forecast financial extremes. Empirical observations of this super-exponential growth before crashes and rebounds in order to predict the turning points have been presented in many papers. For example, the 2006-2008 oil bubble \cite{oil},
the Chinese index bubble in 2009 \cite{Jiangjebo}, the real estate market in
Las Vegas \cite{ZhouSorrealest08}, the U.K. and U.S. real estate bubbles
\cite{Zhou-Sornette-2003a-PA,Zhou-Sornette-2006b-PA}, the Nikkei index
anti-bubble in 1990-1998 \cite{JohansenSorJapan99}, the S\&P 500 index
anti-bubble in 2000-2003 \cite{SorZhoudeeper02}, the South
African stock market bubble \cite{ZhouSorrSA09} and the US repurchase
agreements market \cite{repo}.

In 2008, Lacasa et.~al \cite{lacasa} presented an algorithm which converts a time series into a network. The constructed network inherits many properties of the time series in its structure. For example, fractal time series are converted into scale-free networks, which confirm that power law degree distributions are related to fractality. This method has been applied to the financial time series, \cite{QianJiangZhou10} as well as the energy dissipation process in turbulence \cite{LiuZhouYuan10}.

In this paper, we combine the two powerful tools above and construct an indicator to measure the magnitude of the super-exponential growth of the stock price, by measuring the degree of the price network, generated from the price time series. By investigating twelve major international stock indices, we show that our indicator has strong power to forecast financial extremes.

\section{Methods}
%
Lacasa et.~al presented an algorithm which converts a time series into a network, called \textbf{visibility algorithm}. The detailed construction method is as follows: consider each pair of points in the time series, note these two points as $(t_i, y_i)$ and $(t_j, y_j)$, where $i<j$. Then if all the data points between them are below the line which connect these two points, make a link between these two points. In other words, there is a link between two points $(t_i, y_i)$ and $(t_j, y_j)$, if and only if:
\begin{equation}
y_k < y_i + \frac{t_k-t_i}{t_j-t_i} (y_j-y_i), \forall i<k<j.
\label{eq:linkvis}
\end{equation}
The intuition of this algorithm is simple. Consider each point of the time series as a wall located at the x-axis value $t$ and with the height of the wall is equal to the y-axis value $y$. Assume someone standing on the top of the wall at $(t_i, y_i)$, if she can see the top of the wall at $(t_j, y_j)$, then we make a link between these two points. Of course, the height of this person is ignored here. 

In this paper, we extend the above construction to a new algorithm which can be called the \textbf{absolute invisibility algorithm}. This is just the opposite of the visibility algorithm. Consider two points $(t_i, y_i)$ and $(t_j, y_j)$, where $i<j$. Then, if all the data points between these two points are above the line which connect them, make a link between these two points:
\begin{equation}
y_k > y_i + \frac{t_k-t_i}{t_j-t_i} (y_j-y_i), \forall i<k<j.
\label{eq:linkabsnonvis}
\end{equation}

\begin{figure}[htbp]
\centering
\includegraphics[width=0.45\textwidth]{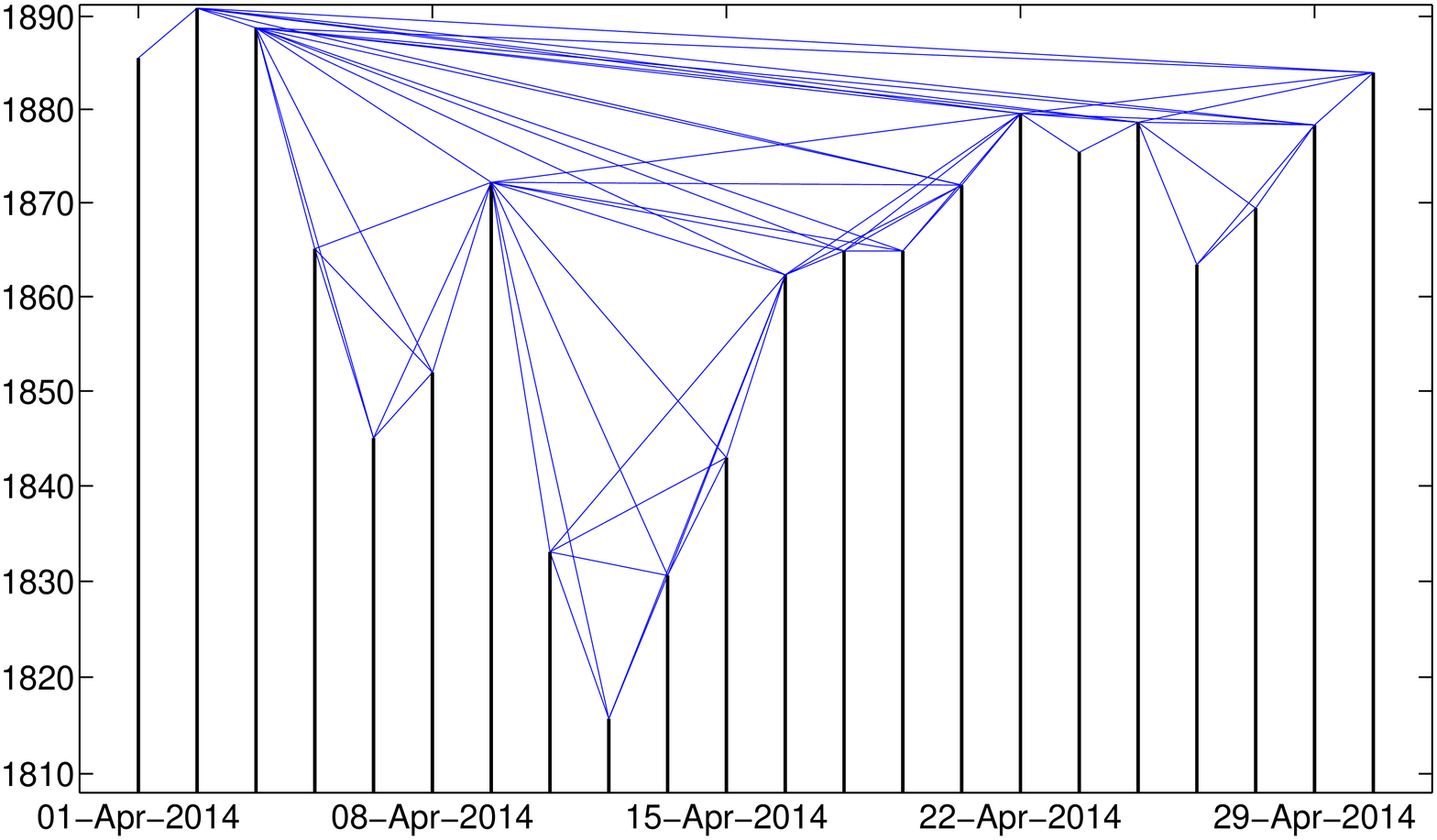}
\includegraphics[width=0.45\textwidth]{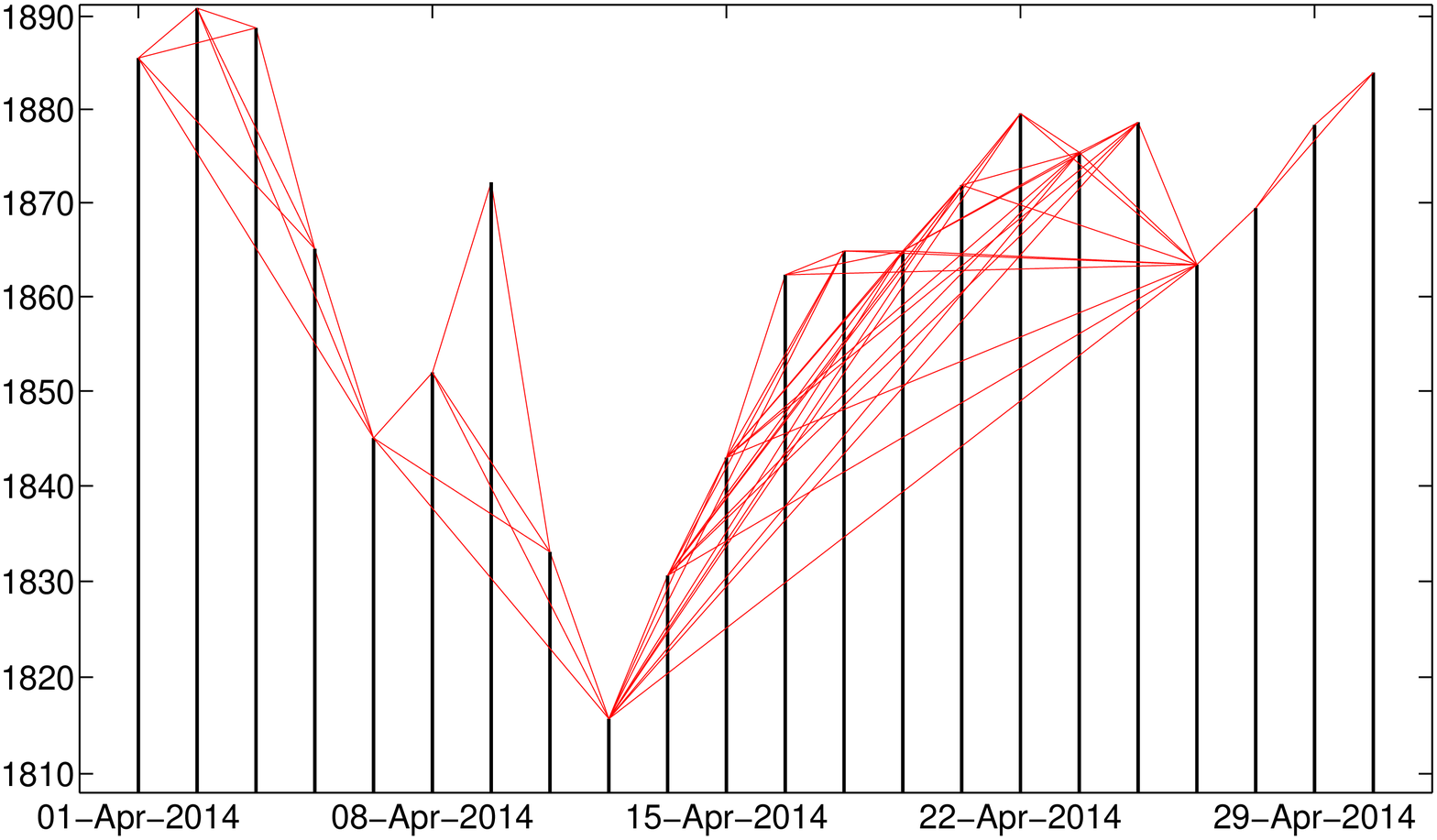}
\caption{Examples of the constructed networks based on the S\&P 500 Index daily close price in April, 2014. Left: the networks constructed by the visibility algorithm. Right: the network built by the absolute invisibility algorithm. Note that the non-trading dates have been removed from the figure.} \label{fig:network_examples}
\end{figure}

To give an idea of what the networks look like in reality, Fig.~\ref{fig:network_examples} shows examples of the constructed networks based on the S\&P 500 Index daily close prices for April, 2014. The upper panel shows the network constructed by the visibility algorithm, while the lower panel shows the network built by the absolute invisibility algorithm. Note that the non-trading dates have been removed from the figures. 

First we start to build the connection between the LPPL model and the visibility/absolute invisibility algorithms. Then, we will explain how to use these two powerful tools to predict financial extremes. 

The LPPL model claims that during the bubble or negative bubble regime, the stock price grows or declines at a super-exponential rate, i.e.~in a bubble regime, both the derivative and the second derivative of the \textbf {log-price} time series are positive; while in a negative bubble regime, both the derivative and the second derivative of the \textbf {log-price} time series are negative. The trick here is to use log-prices instead of normal prices. We know that an exponentially growing time series in a logarithmic-linear scale is represented as a straight line. Therefore, a super-exponentially increasing time series in a log-linear scale is represented as a \textbf{convex} line with a positive slope. In contrast, a super-exponentially decreasing time series in a log-linear scale is represented as a \textbf{concave} line with a negative slope. 

At this point, it is easy to find the connection between the LPPL model and the visibility/absolute invisibility algorithms. The networks, which are constructed by the visibility/absolute invisibility algorithms, contain the information about the super-exponential growth of a time series. In more detail, suppose that an observer stands on the point $(t_i, y_i)$ in a log-linear scaled wall array, if he or she has the visibility of another point $(t_j, y_j)$ from here, we can roughly conclude that the growth rate between these two points is super-exponential. The more points that $(t_i, y_i)$ can see, the higher the confidence that the time series grows in a super-exponential rate at $(t_i, y_i)$. The number of the points that $(t_i, y_i)$ can see is exactly the degree of the visibility network at point $(t_i, y_i)$!

Based on the LPPL model, herding effect and imitation behavior generates super-exponential growth in stock market prices. The fast growth makes the financial system unstable and eventually leads to a sudden regime change, which is often, but not always, a crash in bubble regimes and rebound in negative bubble regimes. In consequence, the confidence of the super-exponential growth rate should be a good proxy to measure how close it is to the point of the regime shift --- the financial extremes. We therefore use the degree of visibility/absolute invisibility network to predict the peaks and troughs in the financial market prices.

However, in order to predict financial extremes, the degree of the visibility/absolute invisibility networks is not sufficient. We still need three more criteria:
\begin{enumerate}
  \item To ensure the \textbf{prediction} condition, for any point in the time series, only the links to its left can be counted. The reason is obvious: at time $t_i$, we do not know the ``future'' information $\{t_j, y_j\}, j = i+1, i+2, \cdots$.
  \item To ensure the \textbf{extreme} condition, the time series should be increasing before the peaks and decreasing before the troughs. Therefore, the link between $(t_i, y_i)$ and $(t_j, y_j)$ (suppose $t_i > t_j$) is made only if $y_i > y_j$ in a visibility network and $y_i < y_j$ in a absolute invisibility network.
  \item To ensure the \textbf{normalization} condition, the same scope should be applied for all the points. There is no link between $(t_i, y_i)$ and $(t_j, y_j)$ if $|i - j| > S$, even though all the other conditions are satisfied.
\end{enumerate}

We build the financial extreme indicators based on the above discussion. A peak degree at time $t_i, i > S$, noted as $D_{peak}(i)$, is defined as the number of $j$s, where $j$ satisfies:
\begin{eqnarray}
  &~&i-S \leq j \leq i-1, j \in \mathbb{N}, \\
  &~&y_i > y_j, \\
  &~&y_k < y_j + \frac{t_k-t_j}{t_i-t_j} (y_i-y_j), \forall j<k<i, \textrm{if} ~ i-j \geq 2.
\end{eqnarray}
By definition, $D_{peak}(i) \in (0, S]$. The peak indicator is defined as $I_{peak}(i) = D_{peak}(i)/S \in (0, 1]$.

Similarly, the trough indicator at time $t_i$ is defined as $I_{trough}(i) = D_{trough}(i)/S$, where  $D_{trough}(i)$, is the number of $j$s, where $j$ satisfies:
\begin{eqnarray}
  &~&i-S \leq j \leq i-1, j \in \mathbb{N}, \\
  &~&y_i < y_j, \\
  &~&y_k > y_j + \frac{t_k-t_j}{t_i-t_j} (y_i-y_j), \forall j<k<i, \textrm{if} ~ i-j \geq 2.
\end{eqnarray}

\section{Results}

We choose twelve international stock indices over 21 years, between Jul. 7, 1993 and Jul. 7, 2014. They are: S\&P 500 composite (S\&PCOMP), Japan TOPIX (TOKYOSE), Hong Kong Hang Seng (HNGKNGI), Russell 1000 (FRUSS1L),	FTSE 100 (FTSE100), STOXX Europe 600 (DJSTOXX), German DAX 40 (DAXINDX), France CAC 40 (FRCAC40), Brazil BOVESPA (BRBOVES), Turkey BIST National 100 (TRKISTB), Mexico Bolsa IPC (MXIPC35) and Thailand Bangkok S. E. T. (BNGKSET). During this period, there are 5479 trading days (note this as $T$). We construct the peak/trough indicators for each of these indices. In this paper, the look-back scope $S$ is set to be 262, since normally one calendar year has about 262 trading days. 

\begin{figure}[htbp]
\centering
\includegraphics[width=0.9\textwidth]{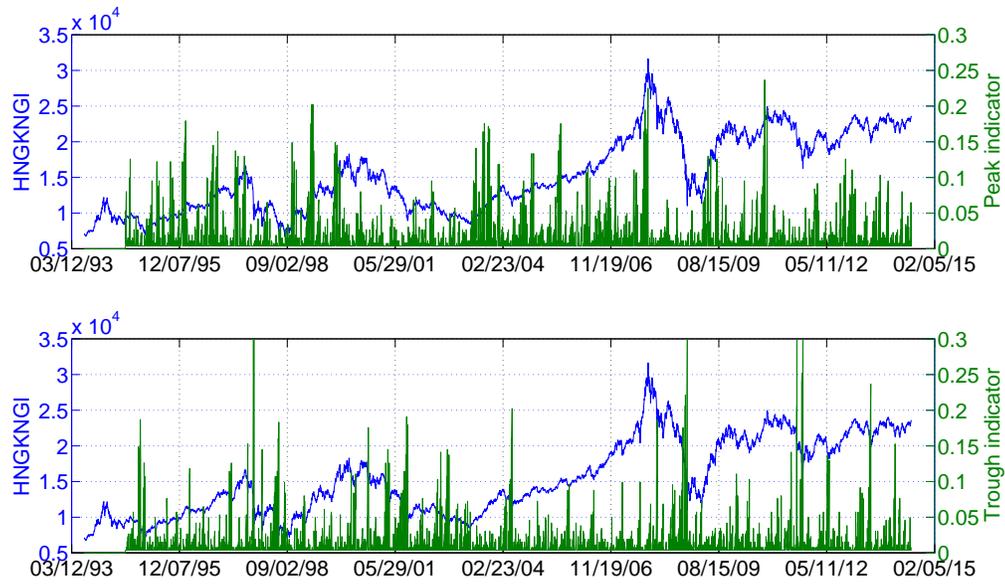}
\caption{Example of the financial extreme indicators as well as the index value based on the Hong Kong Hang Seng Index daily close price for 20 years from Jul. 8, 1994 to Jul. 7, 2014. Upper: the peak indicator. Lower: the trough indicator.} \label{fig:indicator_examples}
\end{figure}

\begin{figure}[htbp]
\centering
\includegraphics[width=0.95\textwidth]{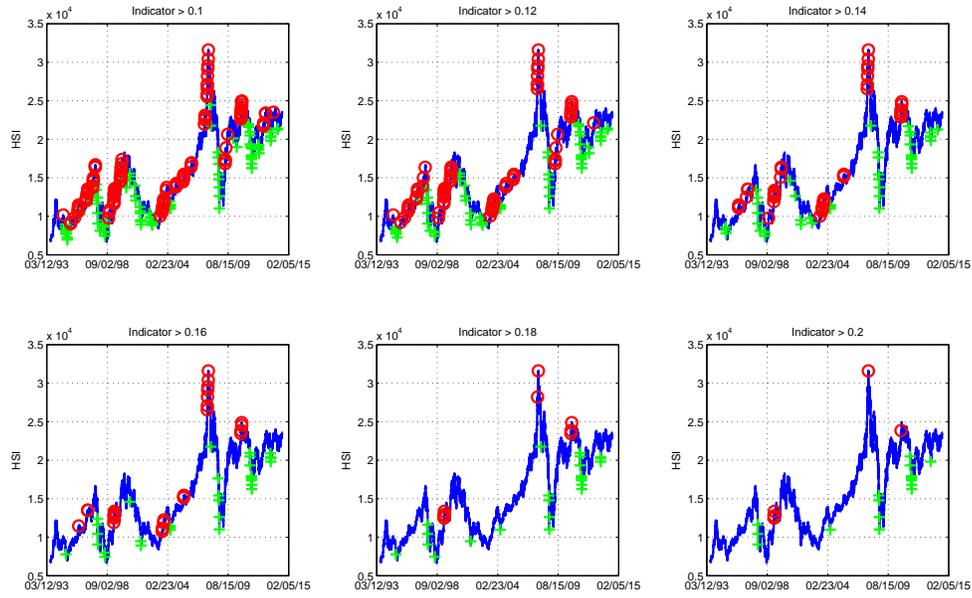}
\caption{Financial extreme indicators with different thresholds. All the points which are marked by red circles are those whose peak indicators are greater than the threshold. All the points which are marked by green crosses are those whose trough indicators are greater than the threshold. As the threshold increases (from 0.1 to 0.2), less predicted peaks and troughs are marked. The sample data is the Hong Kong Hang Seng Index daily close price for 20 years from Jul. 8, 1994 to Jul. 7, 2014.} \label{fig:indicator_threshold}
\end{figure}

We take Hong Kong Hang Seng Index (HSI) as an example to give a general idea on how the indicators look like. Fig.~\ref{fig:indicator_examples} shows the peak and trough indicators of HSI between Jul. 8, 1994 and Jul. 7, 2014 (due to the construction, as the look-back scope $S=262$, the indicators of the first 262 data points are not available). From the figure, one can easily find that when the peak indicator is high, the stock index is usually very close to a local peak; and when the trough indicator is high, the stock index is usually very close to a local trough. To better present this view, Fig.~\ref{fig:indicator_threshold} is a summary of indicated peaks and troughs where the peak/trough indicators are greater than a threshold. By increasing the threshold, there are less and less indicated peaks and troughs. Most of them coincide with the real local peaks and troughs. It confirms that our new indicators have predictive power on real financial extremes.

Fig.~\ref{fig:indicator_examples} and Fig.~\ref{fig:indicator_threshold} show the predictive power of the indicators qualitatively. We also present the predictability of the indicators quantitatively by introducing the error diagram method \cite{Molchan1,Molchan2}.

In order to implement the error diagram, we have to define peaks/troughs first. For one stock index, the peaks/troughs are defined as the points which have the maximum/minimum value in $b$ days before and $a$ days after. Here we use $b=131$ and $a=45$, which correspond to half a year and two months in trading days respectively.
\begin{eqnarray}
  PK &=& \{(t_i,y_i)\}, \textrm{where} ~ y_i = \max(y_j), i-b \leq j \leq i+a, j \in \mathbb{N}, \label{eq:peakdefine}\\
  TR &=& \{(t_i,y_i)\}, \textrm{where} ~ y_i = \min(y_j), i-b \leq j \leq i+a, j \in \mathbb{N}. \label{eq:troughdefine}
\end{eqnarray}

\begin{figure}[htbp]
\centering
\includegraphics[width=0.45\textwidth]{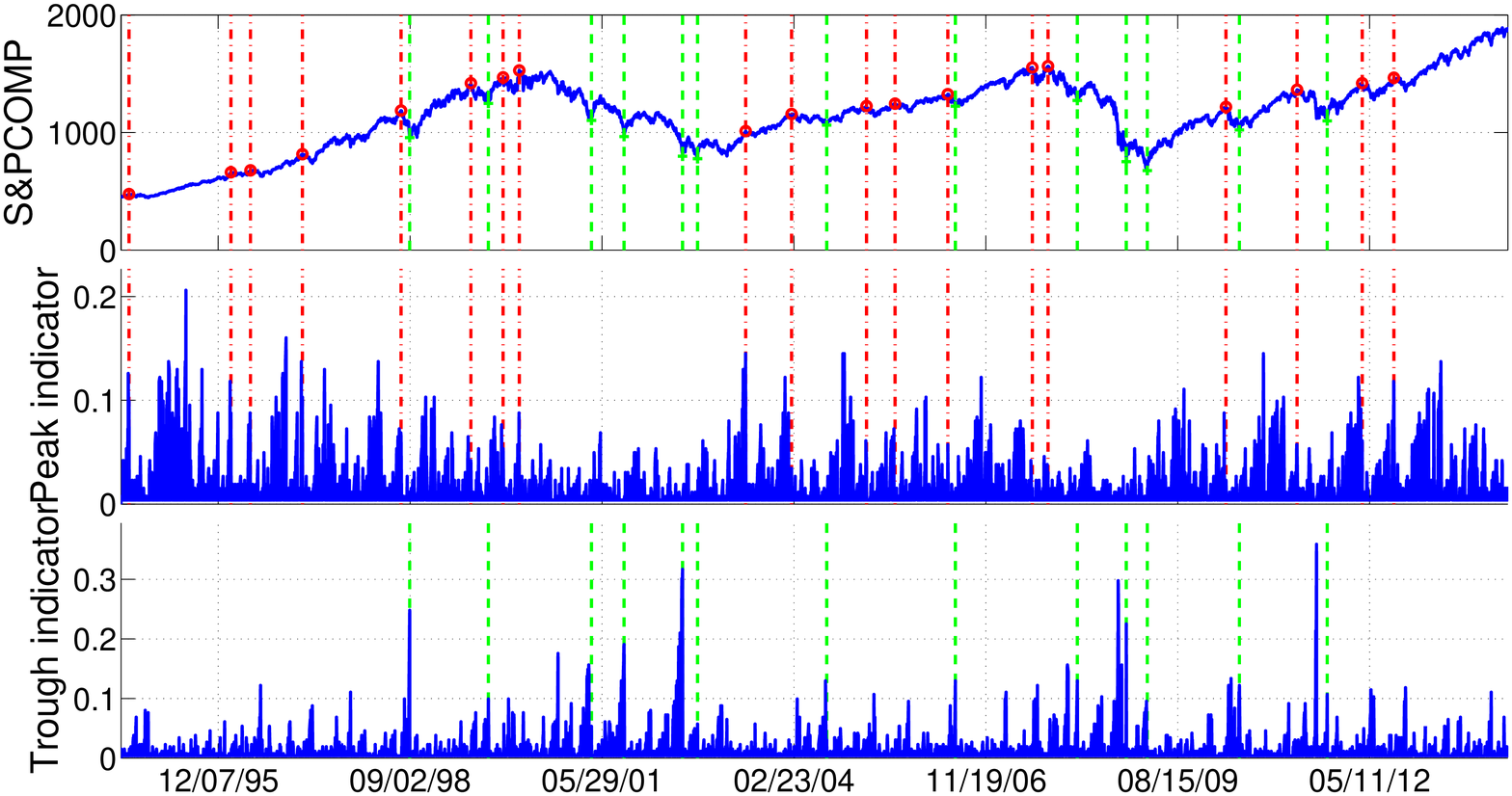}
\includegraphics[width=0.45\textwidth]{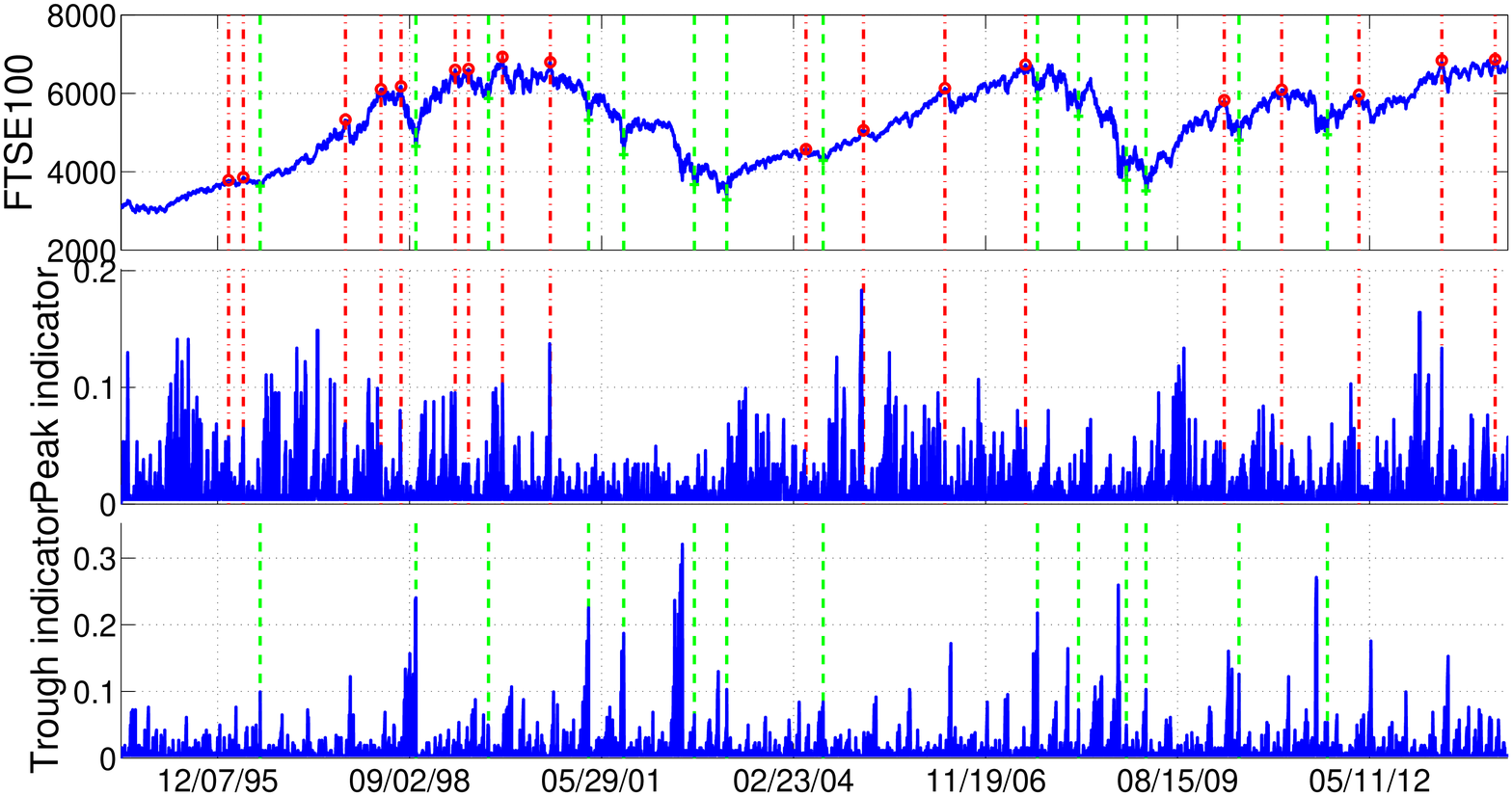}
\includegraphics[width=0.45\textwidth]{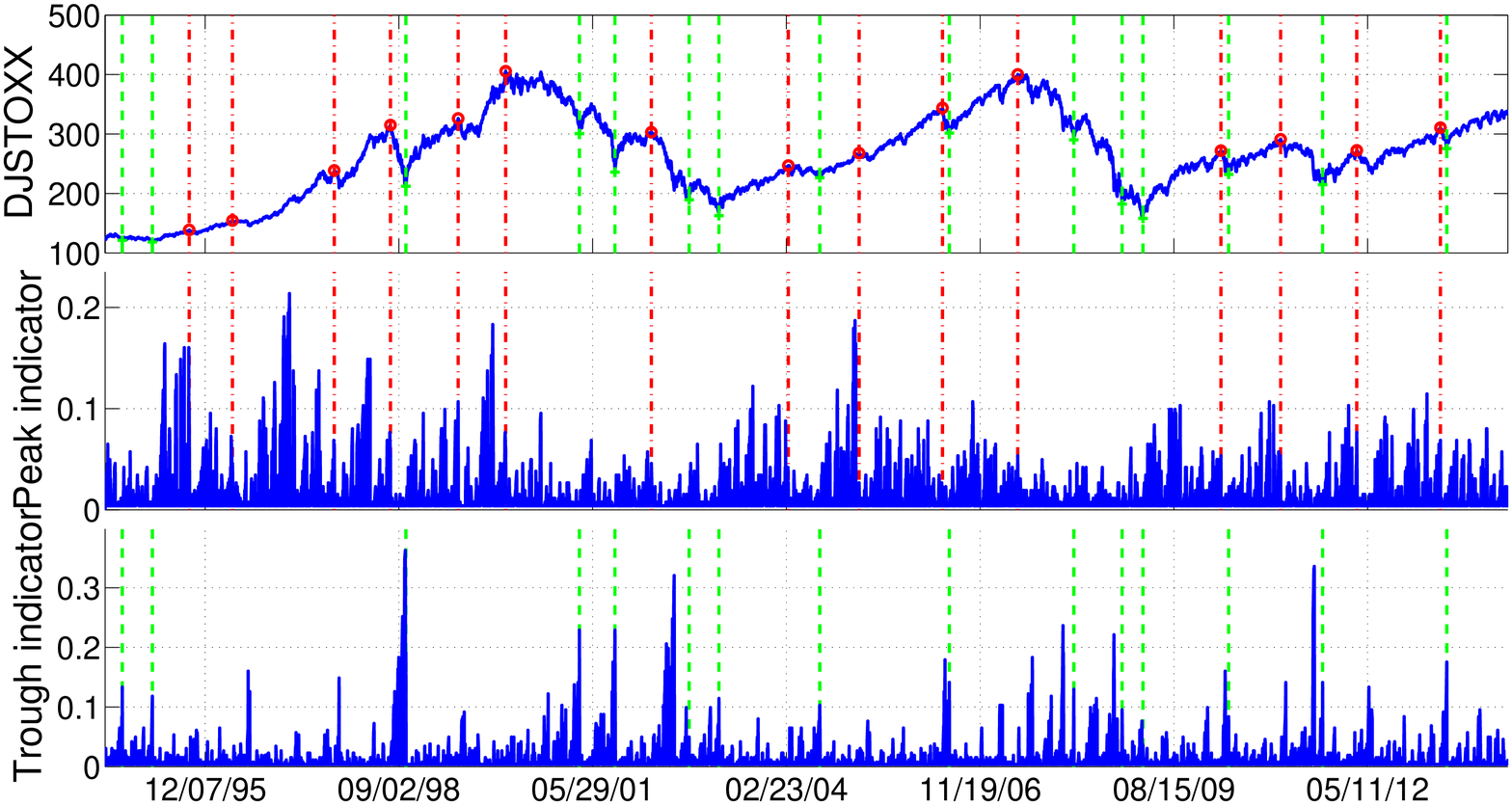}
\includegraphics[width=0.45\textwidth]{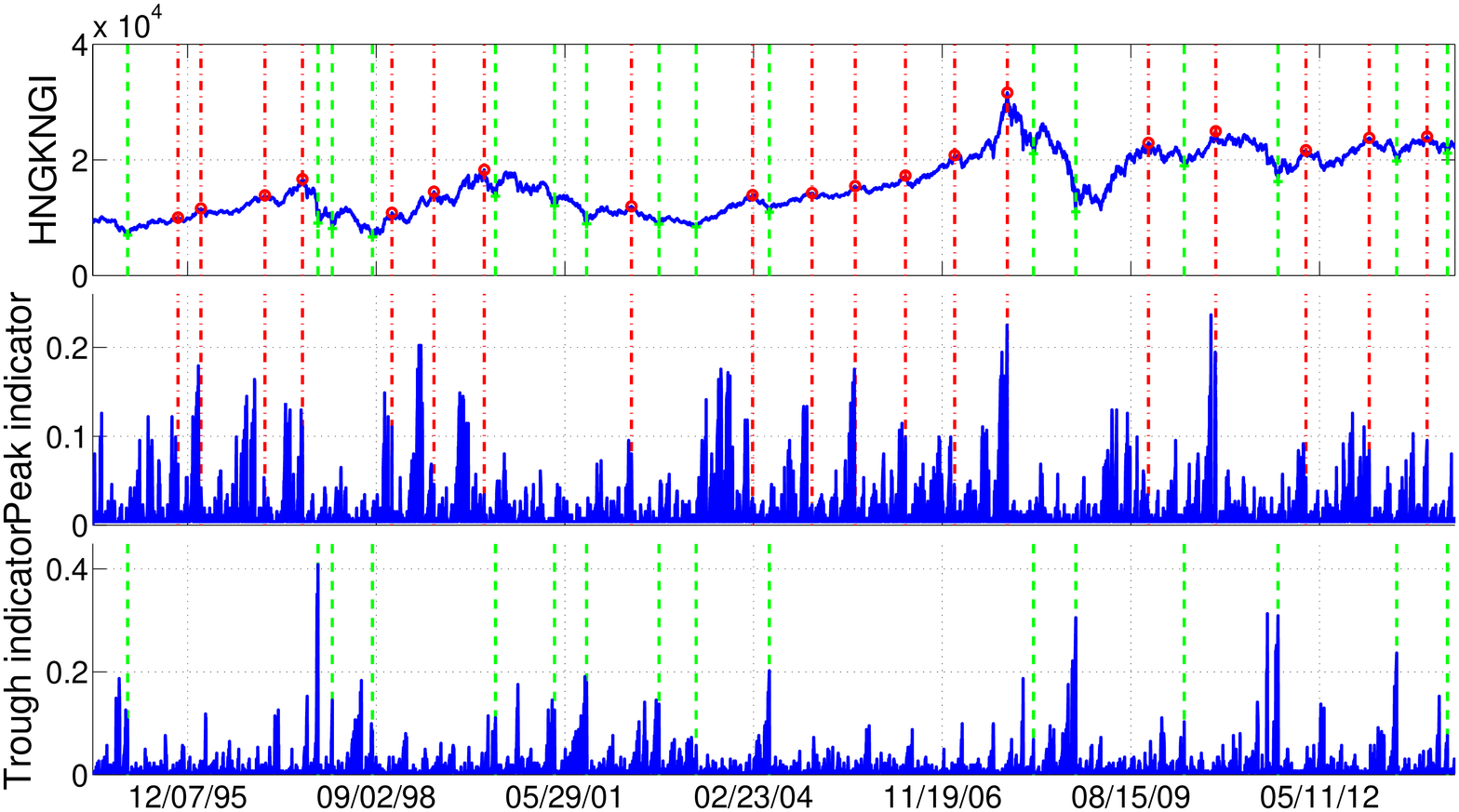}
\caption{Peaks and troughs defined by (\ref{eq:peakdefine}) and (\ref{eq:troughdefine}) together with the peak/trough indicators. Four examples are S\&P 500, FTSE 100, STOXX Europe 600, and Hong Kong Hang Seng. The peaks and troughs are marked by red circles and green crosses respectively. The time when peaks and troughs occurred are also indicated in vertical dash-dotted lines and dashed lines respectively. The sample data are the daily close price of these four indices for 20 years from Jul. 8, 1994 to Jul. 7, 2014.} \label{fig:peak_trough_indicator}
\end{figure}

To give a brief idea that how the peaks and troughs look like in reality, we show the four most important examples: S\&P 500, FTSE 100, STOXX Europe 600, and Hong Kong Hang Seng. In Fig.~\ref{fig:peak_trough_indicator}, we display the index values together with the peak/trough indicators. At the same time the peaks and troughs defined by (\ref{eq:peakdefine}) and (\ref{eq:troughdefine}) are marked by red circles and green crosses respectively. The time when peaks and troughs occurred are also indicated in vertical dash-dotted lines and dashed lines respectively.

An error diagram for predicting peaks/troughs of a stock index is created as follows:
\begin{enumerate}
\item Count the number of peaks defined as expression (\ref{eq:peakdefine}) or troughs defined as expression (\ref{eq:troughdefine}).
\item Take the peak/trough indicator time series and sort the set of all indicator values in decreasing order. Consider that we have to use the first look-back scope days $S$ to generate the indicator and $a$ days in the end where we cannot define peaks/troughs (since we need $a$ days in the future to confirm whether today is a peak/trough or not). The indicator time series here we used is actually $I_{peak}(i)$ and $I_{trough}(i)$, where $S+1 \leq i \leq T-a$.
\item  The largest value of this sorted series defines the first threshold.
\item Using this threshold, we declare that an alarm if the peak/trough indicator time series exceeds this threshold. Then, a prediction is deemed successful when a peak/trough falls inside the alarm period.
\item If there are no successful predictions at this threshold, move the threshold down to the next value in the sorted series of indicator.
\item Once a peak/trough is predicted with a new value of the threshold, count the ratio of unpredicted peaks/troughs (unpredicted peaks (troughs) / total peaks (troughs) in set) and the ratio of alarms used (duration of alarm period / prediction days, where prediction days equals to $T-a-S$). Mark this as a single point in the error diagram.
\item In this way, we continue until all the peaks/troughs are predicted.
\end{enumerate}

The aim of using such an error diagram in general is to show that a given prediction scheme performs better than random.  A random prediction follows the line $y = 1 - x$ in the error diagram. This is simple because that if there is no correlation between the fraction of the predicted events should be proportional to the fraction of the alarms activated during the whole period. A set of points below this line indicates that the prediction is better than randomly choosing alarms. The prediction is seen to improve as more error diagram points are found near the origin $(0, 0)$. The advantage of error diagrams is to avoid discussing how different observers would rate the quality of predictions in terms of the relative importance of avoiding the occurrence of false positive alarms and of false negative missed peaks/troughs. By presenting the full error diagram, we thus sample all possible preferences and the unique criterion is that the error diagram curve be shown to be statistically significantly below the anti-diagonal $y = 1 - x$.

\begin{figure}[htbp]
\centering
\includegraphics[width=0.45\textwidth]{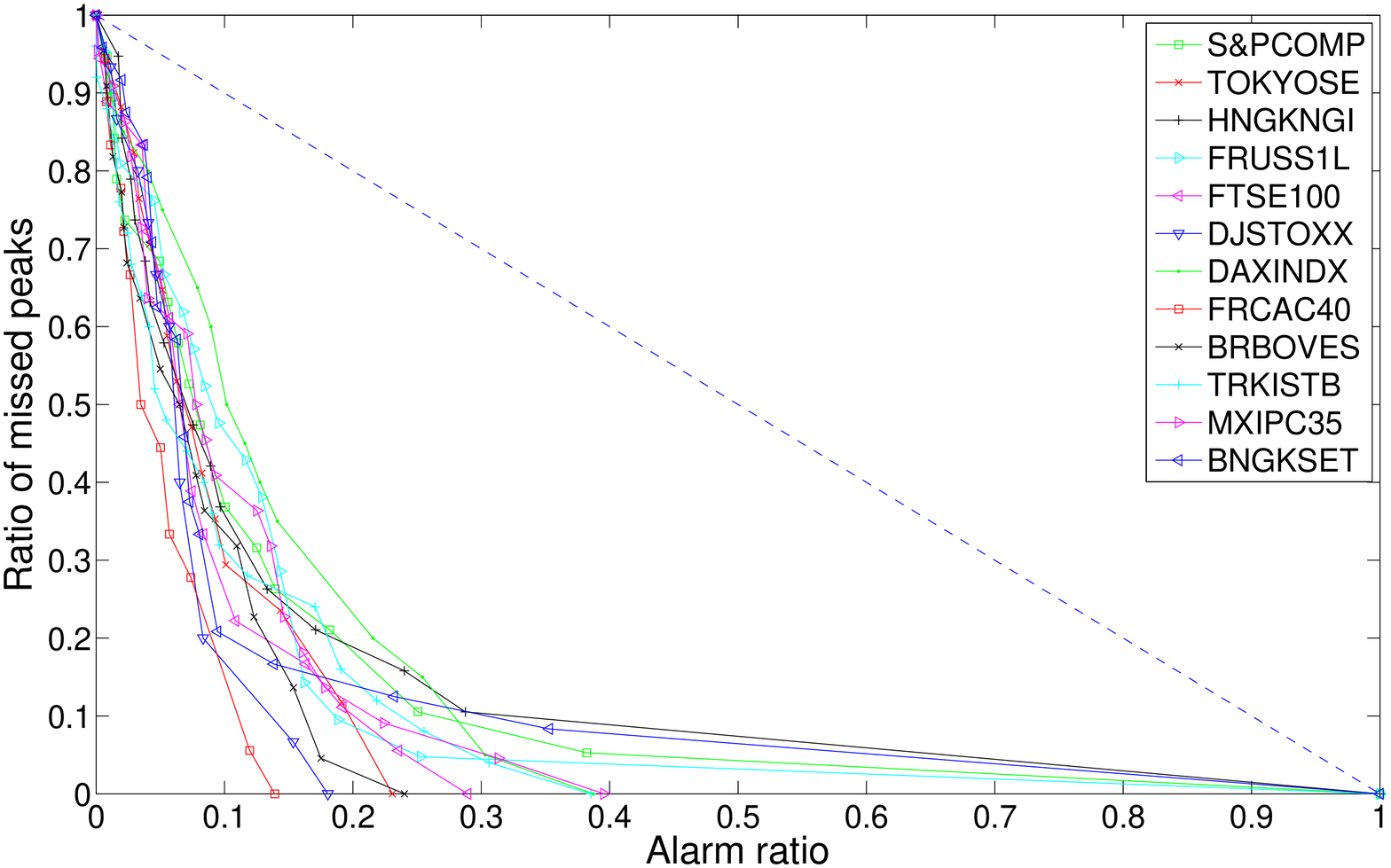}
\includegraphics[width=0.45\textwidth]{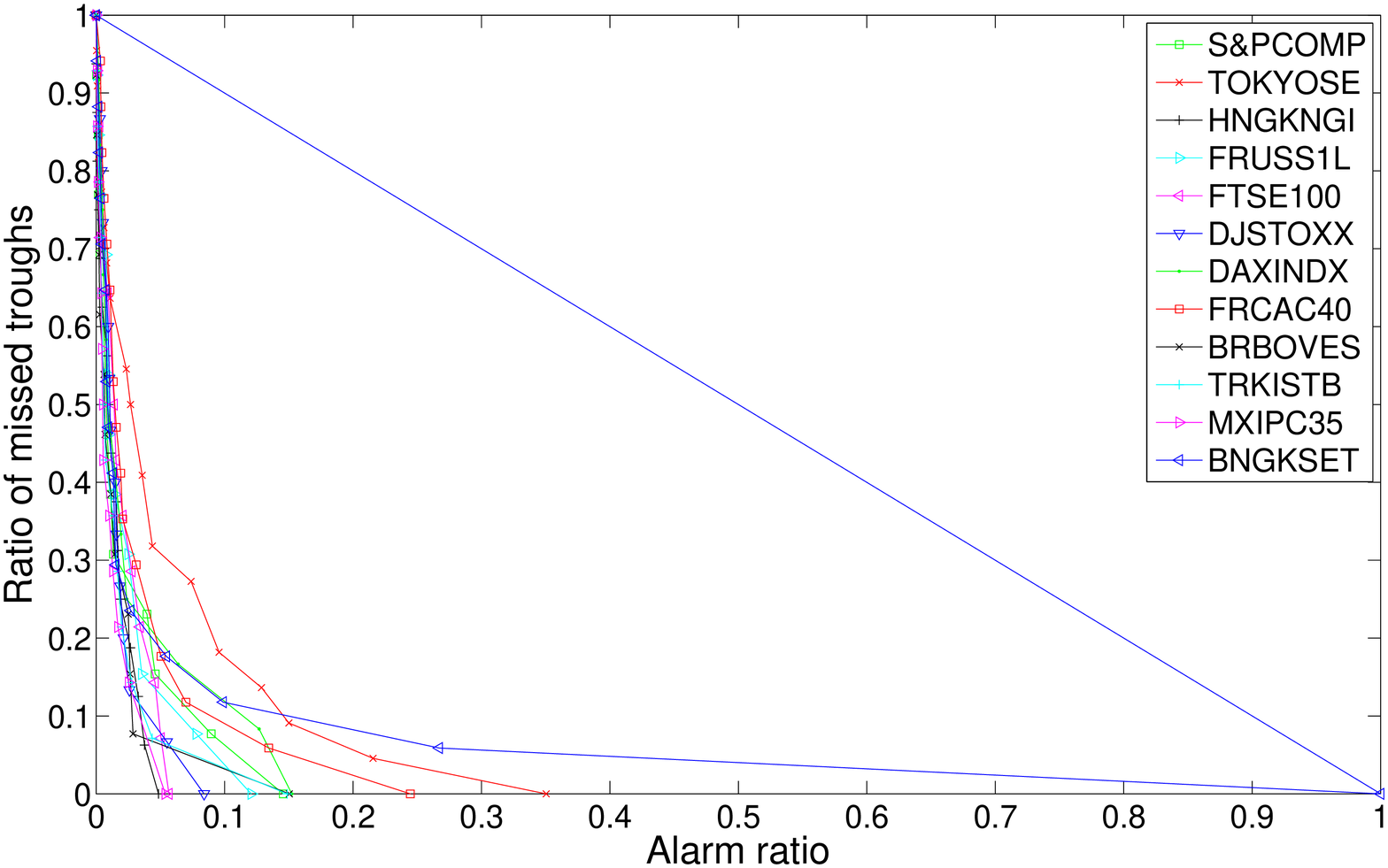}
\caption{Error diagrams for twelve world major stock indices. Left: the error diagram curves for peaks. Right: the error diagram curves for troughs. The line $y=1-x$ represents the random prediction, the more the curves close to the origin $(0,0)$, the better the predictability.} \label{fig:error_diagram}
\end{figure}

In Fig.~\ref{fig:error_diagram}, we show error diagrams for different stock indices. The left panel shows the error diagram curves for peaks, while the right panel shows the error diagram curves for troughs. It is very clear that all these curve, both for peaks and troughs, are far below the random prediction line $y = 1-x$. This means our indicators are very powerful in predicting financial extremes. In general, troughs are better predicted than peaks. This might be due to the fact that stock moves are often wilder when investors panic.

An error diagram curve always starts at the point $(0, 1)$ and ends at the point $(1, 0)$. In between, a random error diagram curve could be any monotone function in the unit square. Therefore, the p-values of our financial extreme indicators should be:
\begin{equation}
  p = A/A_{unit} = A, 
\label{eq:pvalue}
\end{equation}
where $A$ is the area which is determined by the error diagram curve of the peak/trough indicator, the x-axis and the y-axis.

Table.~\ref{tb:pvalue} shows the p-values of peak/trough predictions for all the tested stock indices. The average p-value for peaks and troughs are $0.0991$ and $0.0262$ respectively. This means that the predictability of our peak/trough indicators are significant in a significance level $10\%$ and $3\%$ respectively.

\begin{figure}[htbp]
\centering
\includegraphics[width=0.9\textwidth]{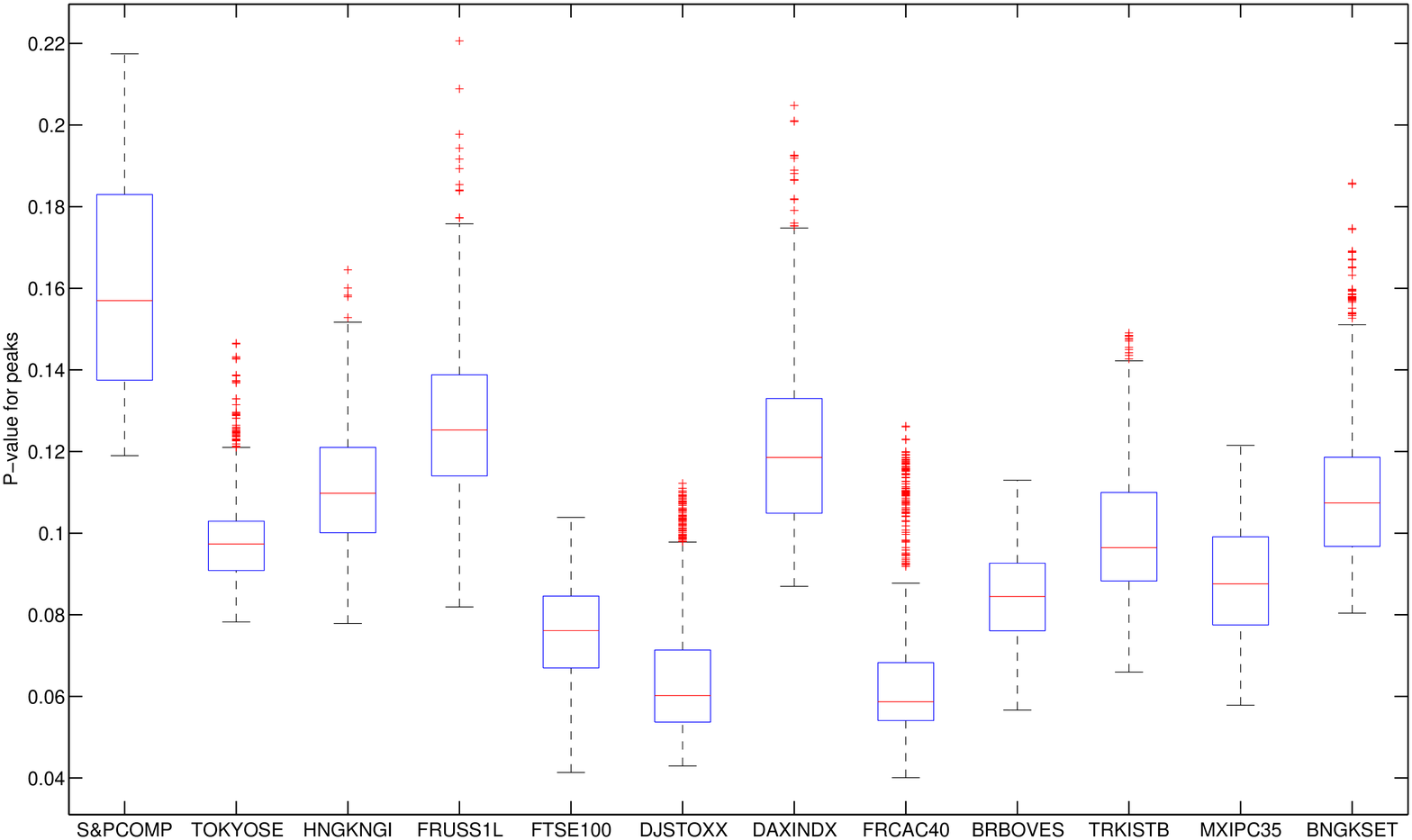}
\includegraphics[width=0.9\textwidth]{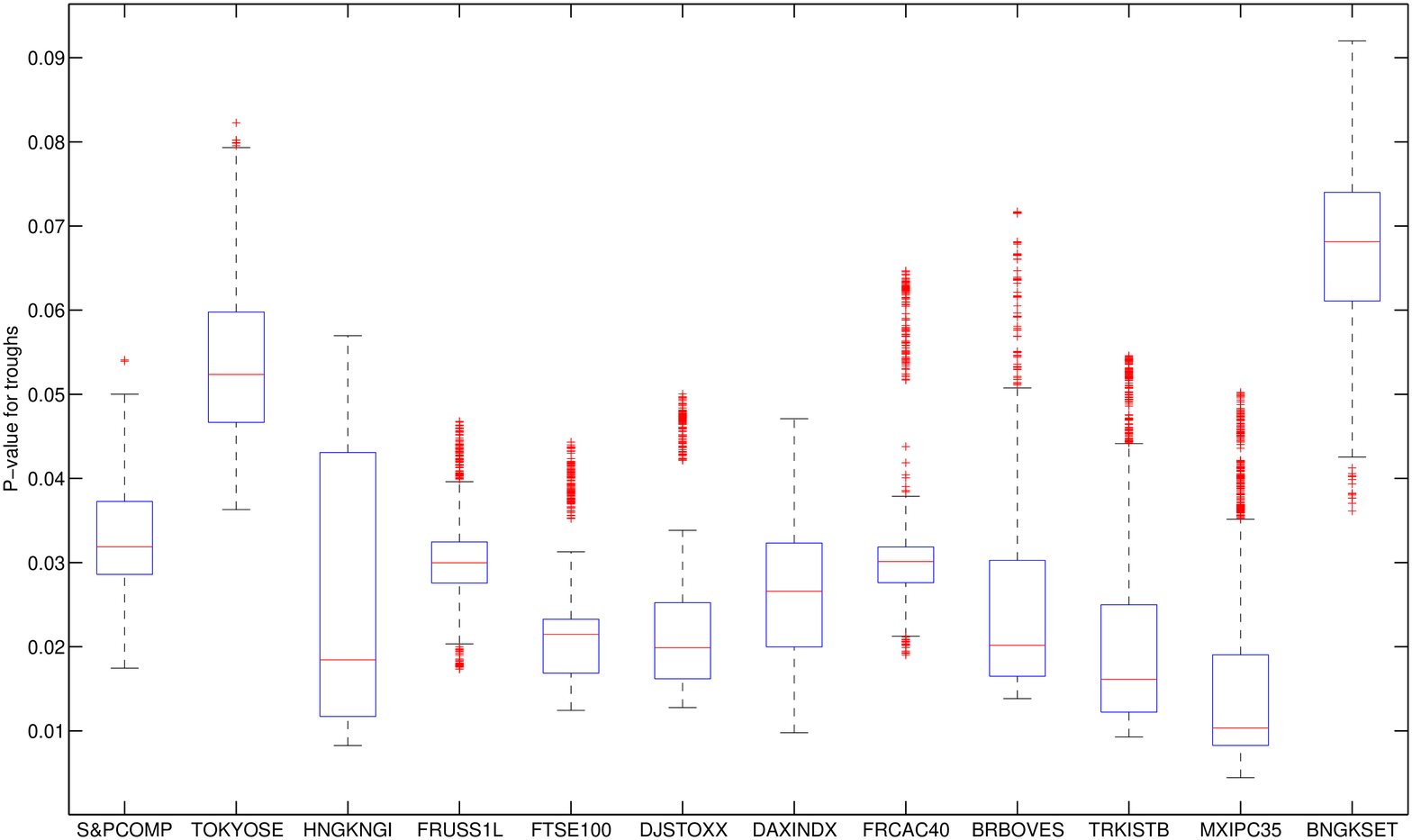}
\caption{Statistics of the p-values of the peak (upper panal) trough (lower panal) predictions. Each box plot represents the p-values of the predictions for 1000 parameter selections of  $S$, $a$ and $b$ defined in (\ref{eq:error_parm}). The lower and upper horizontal edges (blue lines) of box represent the first and third quartiles. The red line in the middle is the median. The lower and upper black lines are the 1.5 interquartile range away from quartiles. Points out of black lines are outliers. The lower the p-value is, the stronger the prediction power.} \label{fig:boxplot_pvalue}
\end{figure}

One may doubt that the selection of the parameters $S$,$a$ and $b$ are arbitrary here. In fact, the selection above is only a reasonable practical decision as $S=262$, $b=131$ and $a=45$ represents the number of trading days of one year, six months and two months respectively. The prediction power of the indicator is generally very strong no matter how we choose the parameters. To prove this, for each stock index, we choose 10 values of $S$, $a$ and $b$:
\begin{equation}
S = 200, 230, 260, \cdots, 470; a,b = 30,45,60, \cdots, 165. \label{eq:error_parm}
\end{equation}
Therefore, we have $10^3=1000$ parameter combinations for each stock index. Fig.~\ref{fig:boxplot_pvalue} shows the statistics of the p-values of the (a) peak and (b) trough predictions. Each box plot represents the 1000 p-values of the predictions: the lower and upper horizontal edges (blue lines) of box represent the first and third quartiles. The red line in the middle is the median. The lower and upper black lines are the 1.5 interquartile range away from quartiles. Points out of black lines are outliers. From the figure, it is clear that no matter how we choose the values of parameters $S$, $a$ and $b$, which market we are measuring, and which type of extremes (peaks or troughs) we are testing, the preditability of our indicator is constantly strong. The mean of the median p-value of peaks and troughs for all the stock indices are 0.0982, 0.0288 respectively. That means  the predictability of our peak/trough indicators are significant in a significance level $10\%$ and $3\%$ respectively.

\section{Discussion}

In this section, we will compare this new indicator with the LPPL pattern recognition indicator. The LPPL pattern recognition indicator is first presented by Sornette and Zhou \cite{SorZhouforecast06}. By introducing the pattern recognition method developed by Gelfand et.~al \cite{gel}, Sornette and Zhou transform the probability prediction of the financial critical time in the standard LPPL model into a quantitative indicator of financial bubbles and crashes. Yan el.~al extended this indicator to financial ``negative bubbles'' and rebounds \cite{rebound}, and made thorough tests on many major stock indices of the world \cite{YanRebWooSor}. Although the p-values of the predictability of this indicator are not clearly calculated in those papers, one can easily make an estimation from the error diagrams in these papers using (\ref{eq:pvalue}). Fig.~5,8 of \cite{SorZhouforecast06} show the predictability of the peaks of Dow Jones Industrial Average Index and Hang Seng Index using the LPPL pattern recognition indicator. Similarly, Fig.~5,7 of \cite{rebound} show the preditability of the troughs of S\&P 500 Index) and Fig.5-7 of \cite{YanRebWooSor} (Fig.~12-14 of the arXiv version) show the predictability of both the peaks and the troughs of Russell 2000 Index, Swiss Market Index and Nikkei 225 Index. For all these examples, the area $A$ determined by the error diagram curve, the x-axis and the y-axis are clearly larger than $0.1$. As $A_{unit} = 1$, the p-values in all these examples are greater than $0.1$. Given that the mean of the median p-value of peaks and troughs for all the stock indices are 0.0982, 0.0288 respectively for our new network indicator, we could conclude that on average the new indicator has a better predictive power than the LPPL pattern recognition indicator.

In summary, we have constructed an indicator of financial extremes via the magnitude of the super-exponential growth of the stock price, by measuring the degree of the price network generated by visibility/absolute non-visibility algorithms with constrains. This new indicator has been applied to twelve major international stock indices. The peaks and troughs of the tested indices over the past 20 years can be effectively predicted by our indicator. The predictability of the indicator has been quantitatively proved by error diagrams. The performance of the indicator is robust to the parameters, and on average better than the LPPL pattern recognition indicator.

\clearpage


\clearpage

\begin{table}[htbp]
\begin{center}
\begin{tabular}{|l|l|l|}
\hline
Ticker&Peak & Trough\\\hline
S\&PCOMP&0.125485&0.023136\\
TOKYOSE&0.087145&0.053484\\
HNGKNGI&0.140173&0.013434\\
FRUSS1L&0.117058&0.021920\\
FTSE100&0.084874&0.019177\\
DJSTOXX&0.065708&0.016427\\
DAXINDX&0.127369&0.029723\\
FRCAC40&0.050226&0.033363\\
BRBOVES&0.074050&0.015495\\
TRKISTB&0.092147&0.016112\\
MXIPC35&0.101478&0.011147\\
BNGKSET&0.123534&0.060996\\\hline
\end{tabular}
\caption{\label{tb:pvalue}P-values of peak/trough predictions for all the tested stock indices.}
\end{center}
\end{table}

\end{document}